# Coulomb drag transistor *via* graphene/MoS$_2$ heterostructures

Youngjo Jin[1,2,†], Min-Kyu Joo[1,2,†,*], Byoung Hee Moon[1], Hyun Kim[1,2], Sanghyup Lee[1,2], Hye Yun Jeong[1], Hyo Yeol Kwak[1,2], and Young Hee Lee[1,2,3,*]

[1]*Center for Integrated Nanostructure Physics (CINAP), Institute for Basic Science (IBS), Suwon 16419, Republic of Korea*

[2]*Department of Energy Science, Sungkyunkwan University (SKKU), Suwon 16419, Republic of Korea*

[3]*Department of Physics, Sungkyunkwan University (SKKU), Suwon 16419, Republic of Korea*

*Correspondence to: mkjoo@skku.edu (M.-K.J.) and leeyoung@skku.edu (Y.H.L.)

[†]These authors equally contributed to this work

Two-dimensional (2D) heterointerfaces often provide extraordinary carrier transport as exemplified by superconductivity or excitonic superfluidity[1,2]. Recently, double-layer graphene separated by few-layered boron nitride demonstrated the Coulomb drag phenomenon: carriers in the active layer drag the carriers in the passive layer[1-3]. Here, we propose a new switching device operating *via* Coulomb drag interaction at a graphene/MoS$_2$ (GM) heterointerface. The ideal van der Waals distance allows strong coupling of the interlayer electron-hole pairs, whose recombination is prevented by the Schottky barrier formed due to charge transfer at the heterointerface. This device exhibits a high carrier mobility (up to ~3,700 cm$^2$V$^{-1}$s$^{-1}$) even at room temperature, while maintaining a high on/off current ratio (~10$^8$), outperforming those of individual layers. In the electron-electron drag regime, graphene-like Shubnikov-de Haas oscillations are observed at low temperatures. Our Coulomb drag transistor could provide a shortcut for the practical application of quantum-mechanical 2D heterostructures at room temperature.

Atomically thin two-dimensional (2D) van der Waals layered materials are environmentally inert with no dangling bonds. This allows for exploring a unique facet of science as well as expanding the realm of practical applications to soft and wearable electronics, with high transparency[4] and stretchability[5,6]. The recently developed wafer-scale 2D materials such as graphene (Gr)[7-9] and transition metal dichalcogenides (TMDs)[10,11], combined with the state-of-the-art top-down technology, further enable the high-density integration of complementary metal-oxide-semiconductor components.

Graphene exhibits high carrier mobilities at room temperature but poor switching capability[7-9,12-14]. Meanwhile, the semiconducting TMD counterparts show high on/off ratios with poor mobilities[10,11,15-20]. Despite tremendous efforts, improvements in the switching capability of Gr[21,22] and the carrier mobilities of TMDs[23-25] have been marginal, and hardly satisfactory for industrial applications. In this work, by constructing $Gr/MoS_2$ heterostructures (GM), we propose a Coulomb drag transistor as a new device platform to realize a high on/off-current ratio ($I_{on}/I_{off}$) of ~$10^8$ and high carrier mobility up to ~3,700 $cm^2V^{-1}s^{-1}$ at room temperature, which has been rarely achieved previously in any other materials or structures.

$MoS_2$ and Gr grown by chemical vapor deposition (CVD) were stacked onto a $SiO_2$ (300 nm)/Si substrate by wet transfer (see Methods). The bottom $MoS_2$ layer was connected to the source, drain, and inner four-probe electrodes for Hall measurements (Figure 1a and b). The top Gr layer was deliberately isolated from all the electrodes. A dual-gate configuration was adopted in our device to control the carrier type and density in each layer. As no buried insulator is used at the heterointerface, a strong interlayer coupling could be established. The monolayer Gr and $MoS_2$ as well as their interlayer interactions were studied by Raman and photoluminescence spectroscopy (see supplementary note 1). Charge transfer occurs from $MoS_2$ to Gr in the GM heterostructure, creating a static dipole layer at the interface. This generates a Schottky (energy) barrier, and consequently prevents interlayer charge recombination (Figure 1c). This energy barrier plays a key role by mimicking a physically isolating insulator, which is essential for stabilizing the conventional Coulomb drag configuration[1-3]. The energy barrier becomes higher when Gr is *p*-type and $MoS_2$ is *n*-type. Figure 1d illustrates the electron-hole (*e-h*) drag configuration in our structure. In long-lived and strongly coupled *e-h* pair states, the holes in *p*-Gr drag electrons in *n*-$MoS_2$ in the GM region.

Figure 2a displays the room-temperature transfer curves as a function of the top-gate bias ($V_{TG}$) at zero back-gate bias ($V_{BG}$). To compare our device characteristics, we fabricated additional individual Gr and $MoS_2$ devices on the same wafer using identical fabrication procedures. The GM device presents an order of magnitude higher $I_{on}/I_{off}$ than the $MoS_2$ device, while maintaining a similar off-current. To clarify this rather unusual carrier transport, four-probe measurements were performed *via* the inner probes of $MoS_2$ (Figure 2b). Longitudinal potential ($V_{XX}$) of the GM device is much smaller than those of $MoS_2$ and Gr under equivalent drain ($V_D$) and $V_{TG}$ biases. Furthermore, a sign reversal of $V_{XX}$ is clearly observed at $V_{TG}$ ~ −13 V, implying *e-h* Coulomb drag behavior which becomes clearer at low temperatures (~ 2K) (see supplementary note 2).

We calculated the four-probe field-effect mobility ($\mu_{FE}$) using the following formula,

$$\mu_{FE} = \frac{L}{W}\frac{g_m}{C_{ox}V_{XX}} \quad (1)$$

where $C_{ox}$, $g_m$, $L$, and $W$ denote the effective oxide capacitance per unit area, transconductance, channel length, and width for the four-probe measurement, respectively[26]. Overall, $\mu_{FE}$ of the GM device exceeds that of $MoS_2$ and even Gr. We define two distinct regimes of the GM device (Figure 2c), based on the charge neutrality point ($V_{CNP}$ = −11 V, white line) of individual Gr; *e-h* and *e-e* drag. In the *e-h* drag regime ($V_{TG} < V_{CNP}$), the divergence of $\mu_{FE}$ is observed, resulting from the nearly zero $V_{XX}$ due to *e-h* drag. Meanwhile, right above $V_{CNP}$, the carrier transport of the GM device is mainly governed by *e-e* drag. At a high $V_{TG}$, far from $V_{CNP}$, the built-in Schottky barrier height is lowered, which allows electrons in $MoS_2$ to transfer to Gr[23]. The carrier transport in this case is dominated by Gr in GM device. The higher $\mu_{FE}$ in this regime is ascribed to the suppression of silicon oxide effects due to screening by underlying $MoS_2$[27,28]. Figure 2d summarizes the excellent performance of our device with remarkably enhanced $\mu_{FE}$ ~3,700 $cm^2V^{-1}s^{-1}$ compared to individual Gr or $MoS_2$, while retaining the high $I_{on}/I_{off}$ of ~$10^8$[7,8,13-15,17,18]. The operation of the GM transistor is governed by the Coulomb drag effect, and is therefore influenced mainly by the interfacial properties, not by the individual layer properties. We also fabricated a series of GM devices with several sets of CVD-grown Gr and $MoS_2$ layers (see supplementary note 3). A similar performance was obtained with both high $I_{on}/I_{off}$ and $\mu_{FE}$, regardless of the sample quality.

To elucidate the main carrier scattering mechanism of the GM device, we measured the temperature (*T*)-dependent transfer curves from 2 to 300K (see supplementary note 4), and plotted the *T*-dependent maximum $\mu_{FE}$ for MoS$_2$, Gr, and GM (Figure 3a), in which the $\mu_{FE}$ values in GM were chosen at $V_{XX}$ ~5 mV below the $V_{CNP}$ to improve the data reliability. Intriguingly, the *T*-dependent mobility of the GM device exceeds those of the individual devices, similar to $V_{TG}$-dependent mobility (Figure 2c). As the temperature decreases from 300 to 100K, the mobility increases by the reduced phonon scattering in the GM and MoS$_2$ devices. Below 100K, the mobility in both devices decreases by the impurity scattering (see supplementary note 4)[19,23]. The mobility in Gr is rather invariant with temperature. The mobility in the GM device is remarkably enhanced at low temperatures, below 50K. Strong *e-h* coupling is formed in this regime, which may induce an alternate transport mechanism, like *e-h* pair condensation at low temperatures.

The strong Coulomb drag effect, however, renders ambiguity in determination of $I_D$ in the MoS$_2$ layer of our GM device, which may consequently lead to over- or under-estimation of $\mu_{FE}$. In addition, the GM device consists of two conducting layers and a vertical tunneling barrier; therefore, the general planar capacitance model may lead to a large discrepancy in the intrinsic mobility. To avoid these ambiguities, the Hall mobility ($\mu_{Hall}$) was determined *via* Hall measurements,

$$\mu_{Hall} = \frac{L}{W}\frac{V_{XY}}{V_{XX}B} \quad (2)$$

where $V_{XY}$ and *B* are the Hall voltage and magnetic field, respectively. $V_{XY}$ and $V_{XX}$ were simultaneously measured in the MoS$_2$ layer for determining $\mu_{Hall}$ as a function of *B*, $V_{TG}$, and $V_{BG}$ at *T* = 2K (Figure 3b). The $V_{TG}$–dependent $\mu_{Hall}$ is almost identical to $\mu_{FE}$ in the *e-h* drag regime (reaching up to ~$10^4$ cm$^2$V$^{-1}$s$^{-1}$). Meanwhile, the $\mu_{FE}$ is underestimated compared to $\mu_{Hall}$ in the *e-e* regime. The mobility divergence is clearly observed at all $V_{BG}$ conditions. The mobility of the GM device near $V_{CNP}$ gradually increases in proportion to $V_{BG}$, owing to strong *e-h* pair coupling.

Another distinct evidence of *e-h* drag is the carrier type conversion in MoS$_2$ of the GM device. The carrier density was extracted from the Hall measurement with *B*-field sweeping (Figure 3c and see supplementary note 5). The sign reversal in the carrier density was clearly observed around $V_{TG} \approx -10$ V, indicating that the electrons in MoS$_2$ are dragged by the holes in Gr.

The anomalously strong Coulomb drag effect was clearly manifested once again under magnetic fields. A signature of the current oscillation in the GM device becomes clear in the $e$-$e$ drag regime, as $V_{BG}$ increased (Figure 4a), owing to strong Coulomb coupling. The Hall resistance ($R_{XY}$, Figure 4b) and magnetoresistance ($R_{XX}$, Figure 4c) at $V_{BG} = 40$ V were monitored in terms of $V_{TG}$. The top-gate bias was intentionally chosen to demonstrate the quantized states ($R_{XY}$ plateaus and $R_{XX}$ minima) near $V_{CNP} \sim -8.2$ V at $V_{BG} = 40$ V (Figure 3b). The quantized $R_{XY}$ values in the plateaus are expressed by the following formula:

$$R_{XY}^{-1} = \pm g\left(n + \frac{1}{2}\right)\frac{e^2}{h} \quad (3)$$

where $g$, $n$, and $h$ are the Landau level (LL) degeneracy, LL index ($> 0$), and Planck constant, respectively[12]. The filling factor ($v$) is then defined as, $v = \pm g(n + 1/2)$. $R_{XY}$ quantization does not perfectly match with the quantum resistance value $(ve^2/h)^{-1}$, because $R_{XY}$ originates from Gr layer but is measured $via$ MoS$_2$ in our device. As illustrated in the inset of Figure 4a, electrons in MoS$_2$ are dragged by Gr electrons and consequently, the quantization of Gr could be mirrored $via$ MoS$_2$. Large resistance fluctuations are observed at $V_{TG} = -8.5$ V due to the unstable states while approaching $V_{CNP}$, even though the quantization occurs at lower LLs. For this reason, we chose $V_{TG} = -7$ V to further discuss the Shubnikov-de Haas oscillation (SdHO) near the $e$-$e$ coupling regime at 2 K and $V_{BG} = 40$ V (see supplementary note 6).

Figure 4d clearly illustrates $R_{XX}$ oscillations and the corresponding $R_{XY}$ plateaus at the filling factors, $v = 14$, 10, and 6, which are consistent with those of Gr[9,12]. The observed SdHOs are persistent even at a relatively high temperature of 100K (Figure 4E). The extracted Berry's phase ($\beta \sim 0.5$, Figure 4f) and cyclotron mass ($m_c \sim 0.03 m_e$, where $m_e$ is the free electron mass, Figure 4g) from $T$-dependent SdHOs, are similar to those of Gr[9,12]. This strongly suggests that SdHOs in the GM device originates from the Gr layer, whereas the MoS$_2$ layer simply projects the SdH oscillations of the Gr layer by the strong Coulomb drag effect.

We designed and demonstrated a device platform of a Coulomb drag transistor as a high performance switching device $via$ Gr/MoS$_2$ heterostructures. The channel carrier mobility could be enhanced using a hexagonal boron nitride gate insulator[27,28] instead of the $Al_2O_3$ top-dielectric which degrades the carrier mobility of the top Gr layer (see supplementary note 3). Several issues remain to be explored toward the practical applications of the Coulomb drag transistor, including contact resistance[23-25], carrier type control[20], and widening of the operation voltage window. Our Coulomb drag transistor opens a new research field for dissipation-less quantum devices with 2D heterostructures operated at room temperature.

**Methods**

**Graphene growth *via* chemical vapor deposition (CVD).** To obtain large single crystalline graphene (Gr) flakes, the following procedure was used. i) A Cu foil was pre-annealed at 1070 °C for 2 h under $H_2$ and Ar atmosphere to remove organic residues from its surface and to improve the crystallinity. ii) One side of the Cu foil was chemically polished *i.e.,* the Cu surface was etched with a Cu etchant ($FeCl_3$, Taekwang, Korea) and rinsed with deionized (DI) water. Thus, a flat and organic residue-free substrate was obtained. Before Gr growth, iii) the Cu foil was annealed again at 1070 °C for 30 min in the growth chamber to remove remaining residues on the surface. iv) A low concentration $CH_4$ gas (0.1% Ar-based gas) was passed. v) After the annealing process, the Cu foil was immediately loaded in the chamber under 3 sccm $CH_4$, 20 sccm $H_2$ (99.9999%), and 1000 sccm Ar (99.9999%) at the same temperature (1070 °C) for 30 min without air exposure, to synthesize high quality Gr flakes.

**$MoS_2$ growth *via* CVD.** We synthesized monolayer $MoS_2$ by sulfurizing a solution-based precursor to improve the crystal quality. The Mo precursor (12 mM ammonium heptamolybdate in DI water (Sigma-Aldrich, 431346)) and a promoter (10 mM sodium hydroxide in DI water (Sigma-Aldrich, 306576)) were mixed and spin-coated onto a $SiO_2$/Si wafer at 3000 rotations per minute (rpm) for 1 min. To sulfurize the Mo-based precursor on the substrate, sulfur powder (100 mg) and the prepared substrate were placed in zone 1 and zone 2 in the growth chamber, respectively, and 500 sccm $N_2$ was passed as a carrier gas during the synthesis. After a sufficient $N_2$ purging process over 1 h, the single-crystalline $MoS_2$ was synthesized for 15 min under atmospheric pressure in different temperature zones; zone 1 (210 °C) and zone 2 (780 °C).

**Device fabrication using the Gr/$MoS_2$ heterostructure (GM)**. To transfer each material, we spin-coated polymethyl methacrylate (PMMA, Sigma-Aldrich) onto the substrate with

Gr/MoS$_2$ at 2000 rpm for 1 min. The flexible PMMA film holds the material and minimizes any damage during the transfer process. The PMMA-coated MoS$_2$ was separated from the substrate by dipping into DI water and transferred onto a SiO$_2$ (300 nm)/Si target substrate. In our MoS$_2$ growth condition, the water-soluble promoter remains near and underneath the MoS$_2$ layer. We removed PMMA by cleaning with acetone, followed by an annealing process at 350 °C for 2 h under H$_2$/Ar atmosphere. The annealing process enhances the interaction between MoS$_2$ and SiO$_2$ as well, resulting in the strong adhesion of MoS$_2$.

To detach PMMA/Gr from the Cu foil, a voltage was applied from a Pt wire to the Cu foil in 0.1 M NaOH solution (water electrolysis method). The chemical doping effect is largely suppressed with this method, which is a key aspect to enhance the device performance. To further remove the chemical residues, Gr was rinsed several times with DI water. We then transferred and stacked Gr onto the MoS$_2$ flakes on the target substrate. The same cleaning and annealing processes (350 °C for 2 h under H$_2$/Ar atmosphere) were conducted to remove residual PMMA and thereby facilitate strong interlayer coupling.

The GM devices were fabricated using standard e-beam lithography and metallization with Cr (5 nm)/Au (60 nm). To define the channel area for the six-probe Hall bar structure, MoS$_2$ and Gr were patterned and etched using O$_2$ and SF$_6$ plasma (10 sccm O$_2$ and 20 W for 20 s, followed by 20 sccm SF$_6$ at 10 W for 10 s). Then, Al$_2$O$_3$ (30 nm) was deposited on the GM devices to serve as a top gate dielectric, *via* atomic layer deposition. The top gate was fabricated by e-beam patterning and Cr (5 nm)/Au (60 nm) deposition to realize a dual-gated device as the final configuration for our experiment.

**Optical characterization.** To identify each material and confirm their interlayer charge transfer effect, Raman and photoluminescence (PL) measurements were performed on GM and MoS$_2$ after Al$_2$O$_3$ passivation. For this, a confocal Raman microscopy (NTEGRA Spectra, NT-MDT) system equipped with a 532 nm excitation laser and an objective lens (numerical aperture of 0.7) was used.

**Electrical characterization.** The device performance was characterized at each fabrication step after annealing at 150 °C for 2 h in high vacuum (~ 10$^{-6}$ Torr). The measurement was performed at room temperature using a standard semiconductor characterization system (4200-SCS, Keithley Instruments). To enhance the data reliability, we also measured the device performance with another characterization system (B1500A, Keysight Technologies).

The temperature- and magnetic field-dependent performances of the devices were monitored under high vacuum (~$10^{-7}$ Torr) in a cryostat (PPMS, Quantum Design, Inc.).

**Supplementary Information** is available in the online version of the paper.

**Acknowledgments** This work was supported by the Institute for Basic Science (IBS-R011-D1), Republic of Korea.

**Author Contributions** Y.J. and M.-K.J. initiated and developed the work under the guidance of Y.H.L. H.K. and S.L. synthesized the monolayer $MoS_2$ and graphene, respectively. Y.J. and H.Y.K. prepared the samples. Y.J. and H.Y.J. measured and analyzed the optical properties of the samples. Y.J., M.-K.J., B.H.M., and H.Y.K. characterized the electrical performance of the devices and performed data analysis. Y.J., M.-K.J., and Y.H.L. wrote the manuscript and discussed with all the authors.

**Author Information** Reprints and permissions information is available at www.nature.com/reprints. The authors declare that they have no competing financial interests. Readers are welcome to comment on the online version of the paper. Correspondence and requests for materials should be addressed to M.-K.J. (mkjoo@skku.edu) and Y.H.L. (leeyoung@skku.edu).

**Figures & Legends**

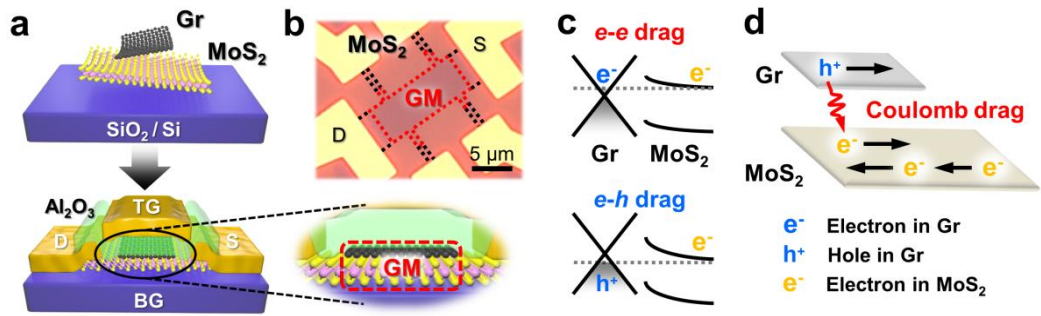

**Figure 1 | Coulomb drag configuration in the Gr/MoS$_2$ (GM) heterointerface. a,** Schematics and **b,** the optical image of the GM device. All the contact electrodes including four-probe contact are connected deliberately to MoS$_2$. **c,** Vertical energy barrier between Gr and MoS$_2$ *via* the formation of a dipole layer created by the charge transfer. The tunneling barrier blocks the additional interlayer charge recombination. **d,** *e-h* Coulomb drag configuration in the GM device. Electrons or holes in Gr drag electrons in MoS$_2$, depending on the gate bias.

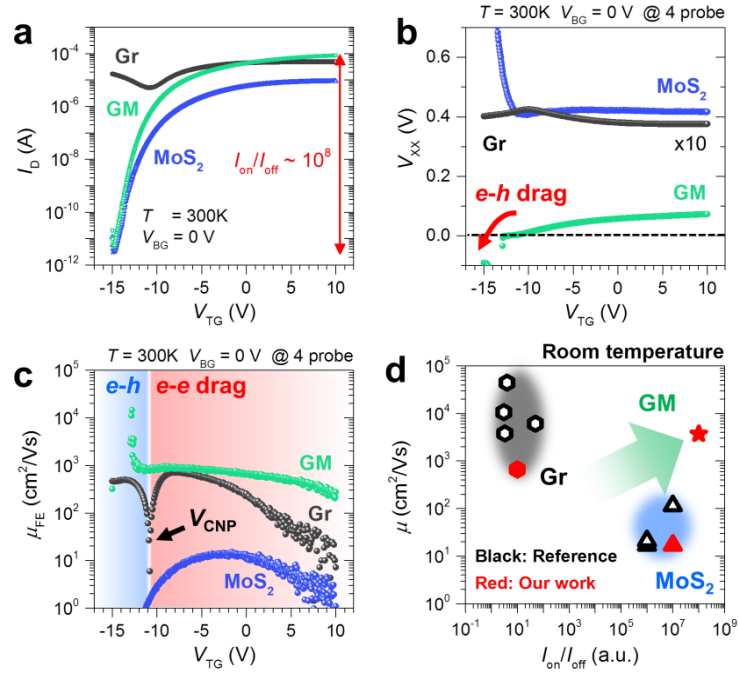

**Figure 2 | Transport characteristics of the Coulomb drag transistor. a,** Room-temperature transfer curves ($I_D$-$V_{TG}$) for the GM device in addition to individual Gr and MoS$_2$ devices for comparison under zero $V_{BG}$. Drain voltage ($V_D$) = 1 V for MoS$_2$ and GM. $V_D$ = 0.1 V for Gr. **b,** $V_{XX}$ curves in each device at 300K and zero $V_{BG}$. For numerical comparison, the Gr signal is normalized by 10 times with respect to the drain voltage. The sign reversal of $V_{XX}$ is clearly observed in the GM device. **c,** $\mu_{FE}$ curve of each device. The Coulomb drag mechanism depends on the carrier type of Gr; $e$–$h$ drag at $V_{TG} \leq V_{CNP}$ and $e$–$e$ drag at $V_{TG} \geq V_{CNP}$. **d,** Comparison of our devices with others in the literature.

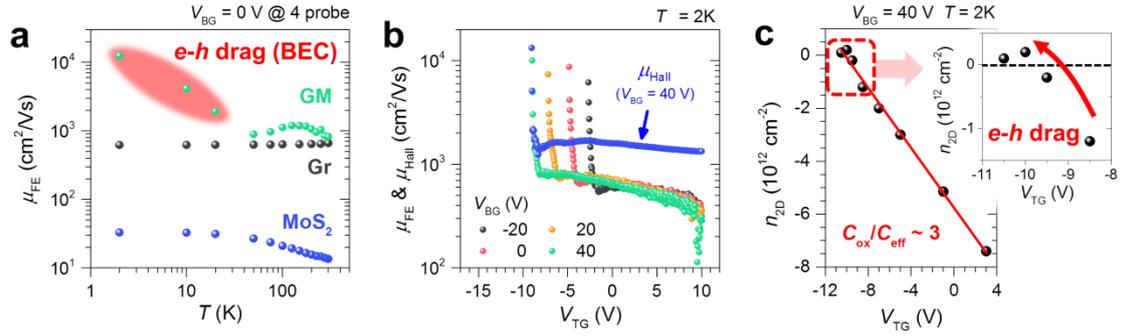

**Figure 3 | T- and B-dependence of the GM Device. a,** $T$-dependent $\mu_{FE}$ of GM, Gr, and MoS$_2$ devices. Mobility of the GM device dramatically increases below $T = 50$K due to strong *e-h* drag. **b,** $V_{TG}$-dependent $\mu_{FE}$ and $\mu_{Hall}$ of the GM device with $V_{BG}$ variance at 2K. **c,** Carrier densities ($n_{2D}$) measured in MoS$_2$ of the GM device at various $V_{TG}$ under 2K and $V_{BG}$ = 40 V. $C_{ox}$ is overestimated by 3 times, compared to the effective capacitance ($C_{eff}$) extracted from the linear fit of $V_{TG}$-$n_{2D}$ obtained from the Hall measurement. Inset: Expanded region of the carrier type conversion. Hole-like carriers are observed in the *e-h* drag regime.

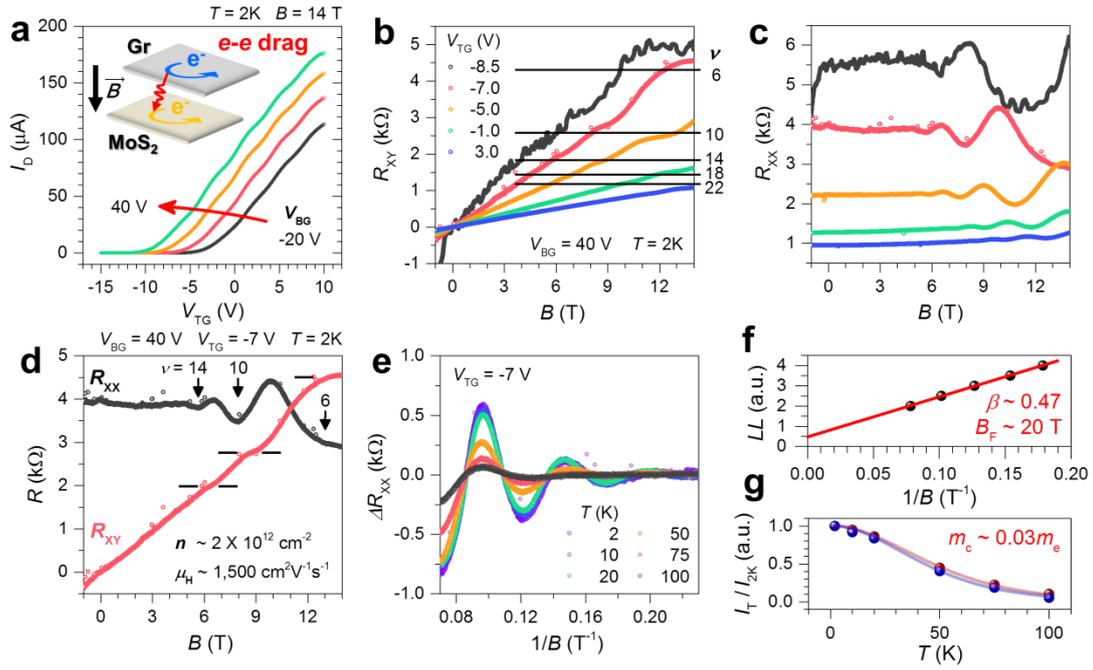

**Figure 4 | Quantized states in GM. a,** $V_{BG}$-dependent transfer curves of GM under $B = 14$ T (normal to the plane) at $T = 2$K. Inset illustrates the *e-e* drag concept. *B*-dependent **b,** Hall resistance ($R_{XY}$) and **c,** magnetoresistance ($R_{XX}$) curves at various $V_{TG}$ under 2K and $V_{BG} = 40$ V. **d,** $R_{XX}$ oscillation and quantized $R_{XY}$ at $V_{TG} = -7$ V and $V_{BG} = 40$ V. **e,** *T*-dependent SdH oscillation. **f,** Berry phase extraction from linear extrapolation (red line). The half Berry phase is the evidence of the Dirac particle of Gr. **g,** The cyclotron mass ($\sim 0.03\ m_e$) extracted from different $R_{XX}$ minima ($\nu = 6$ (red) and 10 (blue)) is close to that of Gr.